\documentclass[
preprint,
pra,
superscriptaddress,
]{revtex4}

\usepackage[mathlines]{lineno}

\usepackage{appendix}
\usepackage{graphicx}
\usepackage{color}
\usepackage{bm}
\usepackage{amsmath, amsthm, amssymb,mathrsfs}
\usepackage{multirow}
\usepackage{todonotes}
\usepackage{csquotes}
\usepackage[colorlinks=true, citecolor=blue,allcolors=blue]{hyperref}
\graphicspath{ {Figs/} } 

\begin{document}

\title{A Multi-Center Quadrature Scheme for the Molecular Continuum}

\author{H. Gharibnejad}
\affiliation{University of Maryland, Department of Chemistry and Biochemistry, College Park, MD, USA }
\thanks{Currently at Computational Physics Inc.}

\author{N. Douguet}
\affiliation{Kennesaw State University, Department of Physics, Marietta, GA, USA}
\affiliation{University of Central Florida, Department of Physics \& CREOL, Orlando, FL, U.SA}

\author{B.~I.~Schneider}
\affiliation{National Institute of Standards and Technology, Applied and Computational Mathematics Division, Gaithersburg, MD, USA}

\author{J. Olsen}
\affiliation{Aarhus University, Department of Chemistry, Aarhus, Denmark}

\author{L. Argenti}
\affiliation{University of Central Florida, Department of Physics \& CREOL, Orlando, FL, U.SA}

\date{\today}

\begin{abstract}
A common way to evaluate electronic integrals for polyatomic molecules is to use Becke's partitioning scheme [{\color{blue}J. Chem. Phys. {\bf 88}, 2547 (1988)}] in conjunction with overlapping grids centered at each atomic site. The Becke scheme was designed for integrands that fall off rapidly at large distances, such as those approximating bound electronic states. When applied to states in the electronic continuum, however, Becke scheme exhibits slow convergence and it is highly redundant. Here, we present a modified version of Becke scheme that is applicable to functions of the electronic continuum, such as those involved in molecular photoionization and electron-molecule scattering, and which ensures convergence and efficiency comparable to those realized in the calculation of bound states.
In this modified scheme, the atomic weights already present in Becke's partition are smoothly switched off within a range of few bond lengths from their respective nuclei, and complemented by an asymptotically unitary weight. The atomic integrals are evaluated on small spherical grids, centered on each atom, with size commensurate to the support of the corresponding atomic weight. The residual integral of the interstitial and long-range region is evaluated with a central master grid. The accuracy of the method is demonstrated by evaluating integrals involving integrands containing Gaussian Type Orbitals and Yukawa potentials, on the atomic sites, as well as spherical Bessel functions centered on the master grid. These functions are representative of those encountered in realistic electron-scattering and photoionization calculations in polyatomic molecules.
\end{abstract} 



\maketitle

\section{INTRODUCTION}
\label{sec:1}

Numerical integration is required for many electronic structure calculations where orbitals other than Gaussians are needed~\cite{Saad06}. Most modern quantum chemistry packages employ Gaussian-type orbitals (GTOs) to describe molecular bound states, a computationally efficient choice, since electronic integrals involving only GTOs may be evaluated analytically~\cite{Alder63,Jeppe2012}. Although optimum GTO basis sets have already been designed  to reproduce highly excited Rydberg states \cite{Kaufmann_1989}, nodeless GTOs are ill-suited to describe continuum functions. Indeed, extended sets of GTOs can reproduce at most a handful of the characteristic radial oscillations of continuum functions before becoming linearly dependent~\cite{Marante2014}.

Modern approaches to represent continuum wave functions in molecular systems rely on a hybrid basis set that comprises both  GTOs, centered on each atom and on the molecular center of gravity, as well as numerical functions such as Bessel functions, Coulomb functions~\cite{Rescigno95a}, B-splines~\cite{Marante2014,Marante2017,Zdenek2020} or finite-element basis functions~\cite{Rescigno05}, to properly describe both the short and long-range behavior of continuum orbitals. Such hybrid approaches allow one to take advantage of the capabilities of modern quantum chemistry packages to compute neutral and ionic bound states but require the electronic integrals between hybrid functions to be evaluated numerically. Numerical quadrature algorithms are often easier to code and parallelize than analytic techniques but can converge slowly due to the sharp variations of molecular orbitals and interaction potentials in the proximity of the atomic nuclei. These limitations are particularly severe in single-center expansion techniques~\cite{BISHOP1967}. To achieve uniform numerical convergence, therefore, it is essential to employ quadrature grids that selectively cluster at all nuclei where the wave function does not vanish.

Several techniques have been developed to evaluate hybrid integrals~\cite{LACBISLA,Gianturco1994,Yip2010,Yip2014}. The simplest approach consists in expanding any polycentric GTO function in terms of a monocentric auxiliary basis~\cite{Zdenek2020}.  As already noted, this approach converges slowly, since representing the sharp variation of a molecular orbital in the neighborhood of an off-center nucleus with acceptable accuracy requires a large set of angular momenta or angular integration points as well as high radial resolution near each nucleus. As a consequence, the auxiliary basis or grid must be chosen to cover the whole molecular region, both close to and far away from the atoms, thus drastically increasing the number of integration points. An alternative approach is to confine the numerical basis to a radial range outside a spherical region that encompasses all the nuclei~\cite{Marante2017}. In this approach, the hybrid integrals are easy to evaluate, but the Gaussian basis must be able to reproduce the continuum orbitals in the whole internal region, thus limiting the size of the molecular system. Neither of these approaches, therefore, offers a fully satisfactory solution to the construction of scattering functions for general molecules. In particular, they are poorly suited to describe molecular dissociation. 

To account for the sharp variation of poly-centric molecular wave functions close to the nuclei, it is possible to separately integrate the region in close proximity of each nucleus. Along this direction, pioneered by Boys and Rajagopal~\cite{Boys1965}, are partitioning schemes that divide the molecular space into Voronoi polyhedra, each containing a single atom~\cite{Boerrigter1988,Averill1989,Pederson1990,TeVelde1992}. The integral within the sphere inscribed in the polyhedron centered on an atom is evaluated using single-center techniques on the atomic site. While the Voronoi method effectively tackles the nuclear region, the evaluation of the residual polyhedron integral outside the sphere is still a challenge, due to the complex shape of the Voronoi boundary. Furthermore, the use of Voronoi polyhedra is still not suited to the electronic continuum.
 
In 1988, Becke has proposed an integration scheme that 
 divides space into overlapping ``fuzzy" Voronoi polyhedra~\cite{Becke1988,BeckeDickson1988}, i.e., a set of smooth positive weight functions $w_\alpha(\vec{r})$, one per nucleus, defining a partition of unity, $1=\sum_\alpha w_\alpha(\vec{r})$, with the additional requirement that each weight function vanishes at all but one of the nuclear positions $R_\beta$, $w_\alpha(\vec{R}_\beta)=\delta_{\alpha\beta}$. Using Becke's partitioning, it is possible to split any electronic integral $I = \int \mathrm{d}^3\vec{r} f(\vec{r})$ into separate atomic components, $I=\sum_\alpha I_\alpha$, $I_\alpha = \int \mathrm{d}^3r w_\alpha(\vec{r})f(\vec{r})$, whose argument is regular everywhere except at a single nucleus, and is accurately discretized with a spherical quadrature grid centered on that nucleus. 
Becke scheme, which is easier to implement than methods based on non-overlapping Voronoi polyhedra, has been demonstrated to be at least as accurate and efficient when applied to density functional calculations~\cite{Franchini2013}, and is now prevalent in the density-functional methods of common quantum chemistry packages~\cite{MOLCAS2003,QCHEM2015,ADF2001,MOLPRO_brief,barca_recent_2020,TURBOMOL2020}. 
Becke scheme is also a promising starting point to evaluate more general multi-center integrals, such as the bound-free and free-free integrals that appear in scattering and photoionization problems~\cite{Rescigno95a}. 

In the original formulation of Becke scheme, the atomic weights do not vanish asymptotically, and hence each atomic grid must cover a region that may be as large as the whole molecular electronic density in the integral. This circumstance limits the applicability of the method in ionization problems, where the electronic density is finite even at a large distance from the molecule and each atomic grid would consequently need to be as large as the whole quantization volume for the continuum. Furthermore, the integral suffers from slow convergence due to the mismatch between the boundary of the quantization volume and that of the atomic spherical grids. 

In this work, we propose an extension of Becke scheme that circumvents this difficulty and which is tailored to treat photoionization of polyatomic systems. This new scheme defines a partition of unity in which the atomic weights $w'_\alpha(\vec{R}_\beta)=\delta_{\alpha\beta}$ are confined to the molecular region, $\sum w_\alpha'(\vec{r})\simeq 0$, $ \forall r>R_{\textsc{Mol}}$, and complemented by a positive external weight, $w_0'(\vec{r})=1-\sum_\alpha w_\alpha'(\vec{r})$, with $w_0'(\vec{R}_\alpha)=0$,  $\forall\alpha$. In this new approach, the atomic grids can be confined within spherical shells of the size of the first one or two nearest-neighbor atoms, whereas the complementary interstitial and asymptotic portion of the integrand are regular everywhere and hence can be discretized with a single master spherical grid, which is typically placed at the center of gravity of the molecule.  The method has all the advantages of the Becke scheme near the nuclei and is capable of treating integrals  which have appreciable contributions far from the atomic centers by employing a single-center master grid at larger distance. As a result, the method is more efficient than the original scheme, and it is flexibile in the choice of the shell's radii. In the mid- to long-range region, where only a limited number of multipolar terms contribute, the central grid can, by itself, describe the numerical integration with a limited number of angular points. The use of Becke weights in the present approach is essential to allow distinct atomic shells to overlap.

In the next section we illustrate the modified Becke schemes, present two variants of the methods, and discuss their accuracy and efficiency for a set of integrals of representative functions. The functions are chosen to represent both typical bound state integrals as well as those involving molecular continuum functions.  In Sec. \ref{sec:3} we present test results for a multi-center model that mimics the NO$_2$ molecule. In Sec. \ref{sec:4} we offer our conclusions.

\section{A modified Becke integration scheme}\label{sec:2}
Let us summarize how the weights in the original Becke scheme are defined~\cite{Becke1988} for a polyatomic system with \(N\) nuclei, centered at \(\vec{R}_a\), \(\vec{R}_b\), etc.
The elliptic coordinate
\begin{equation}
\mu_{ab}(\vec{r})=\frac{r_a-r_b}{R_{ab}},
\end{equation}
where $r_a=|\vec{r}-\vec{R}_a|$ and $R_{ab}=|\vec{R}_a-\vec{R}_b|$, 
quantifies how close a position vector $\vec{r}$ is to one or the other of two nuclei, $\vec{R}_a$ and $\vec{R}_b$. The quantity $\mu_{ab}(\vec{r})$ is bounded by $+1$, along the semi-axis $\ell_{ab}=\{(1+x)\vec{R}_{a}-\vec{R}_b, x>0\}$, on the side of atom $a$, and $-1$, along the semi-axis $\ell_{ba}$. The level surfaces of $\mu_{ab}(\vec{r})=c$ are hyperboloids. The plane $\mu_{ab}(\vec{r})=0$ orthogonally bisects the segment  connecting the two nuclei. Consider now a non-increasing positive step function $s(\mu): [-1,1]\to[0,1]$ that transitions from $s(-1)=1$ to $s(1)=0$ and is flat at the two ends of the interval, $\mathrm{d}s/\mathrm{d}\mu|_{\pm1}=0$. The function $s[\mu_{ab}(\vec{r})]$ is then unity for $\vec{r}=\vec{R}_a$ and zero for $\vec{r}=\vec{R}_b$. To define the domain of any atom $a$, therefore, one considers the product of such switch-off functions for all the other atoms in the molecule,
\begin{equation}
P_a(\vec{r})=\prod_{b\neq a}s[\mu_{ab}(\vec{r})],\qquad P_a(\vec{R}_{b})=\delta_{ab}.
\end{equation}
Finally, from the $P_a(\vec{r})$, a partition of unity $\{w_a(\vec{r})\}$ is readily obtained as
\begin{equation}
w_a(\vec{r})=P_a(\vec{r})/\sum_{b}P_b(\vec{r}),\quad \sum_{a}w_a(\vec{r})=1.
\end{equation}
Becke suggested a hierarchy of increasingly sharper polynomial step functions, $s_k(\mu)=[1-f_k(\mu)]/2$, $k\in\mathbb{N}$~\cite{Becke1988}, with 
\begin{equation}
f_1(\mu)= (3\mu-\mu^3)/2,\quad f_{k+1}(\mu)=f_k[f_1(\mu)],
\end{equation}
which we will also use in the modified Becke scheme.

Let's now consider a hybrid basis set composed of GTOs and numerical functions. The matrix element of a local one-body operator $\hat{o}$ between a molecular orbital $\phi(\vec{r})$ and a numerical function $\chi(\vec{r})$ can be partitioned as

\begin{equation}
\begin{split}
\langle \phi | \hat{o} |\chi\rangle &= \sum_{a} \langle \phi | w_a\hat{o} |\chi\rangle =\\
&=\sum_{a} \int w_a(\vec{r})\, \mathrm{d}^3 r\,\phi^*(\vec{r})\hat{o}\,\chi(\vec{r})\,.
\end{split}
\end{equation}
Due to the presence of the weighting factor $w_a(\vec{r})$, the argument of this integral smoothly approaches zero at every nucleus other than $a$. It is then natural to discretize each atomic integral $\langle \phi | w_a\hat{o} |\chi\rangle$ in terms of a sum over the points of a spherical grid centered on $a$. The atomic integrals are expected to converge faster with respect to the number of angular and radial points of the corresponding atomic grid, which clusters at $\vec{R}_a$, than if a single spherical grid was used for all the atoms in the molecule. As mentioned in the introduction, this considerable improvement in convergence comes at the cost of summing over the points of $N$ grids, instead of a single one. Furthermore, the weight functions introduce an additional modulation of the integrand, which can also affect convergence. 

Let's examine how these ideas impact the calculation of the two-body integral
\begin{equation}
 [\phi_1\chi_1|\phi_2\chi_2] = \int \mathrm{d}^3r \int \mathrm{d}^3 r' \frac{\rho_1(\vec{r}')\rho_2(\vec{r})}{|\vec{r}-\vec{r}'|},\quad\rho_i = \phi_i\chi_i,
\end{equation}
where we assume for simplicity that all the functions, and with them the charge densities, are real. This problem is normally divided into two parts: i) the solution of the Poisson equation for one distribution of charge, say $\rho_1$, 
\begin{equation}
\Phi_1 (\vec{r})= \int \mathrm{d}^3 r' \frac{\rho_1(\vec{r}')}{|\vec{r}-\vec{r}'|},
\end{equation}
or, equivalently,
\begin{equation}
\nabla^2\Phi_1 =-4\pi\rho_1\quad{\rm with}\quad\lim_{r\to\infty}\Phi_1(\vec{r})=0,
\end{equation}
and ii) the subsequent evaluation of the electrostatic energy of the second charge in the field of the first,
\begin{equation}
 [\rho_1|\rho_2] = \int \mathrm{d}^3 r\,\,  \Phi_{1}(\vec{r})\rho_2(\vec{r}).
\end{equation}
In this case, the single-center expansion is dramatically inadequate. Imagine computing  the electrostatic repulsion between two distributions of charge with an appreciable concentration near the same off-center nucleus. The integral between the potential and the charge density has the same poor convergence as in the one-body case.  Worse still, the solution of the Poisson equation would be drastically inaccurate in proximity of the off-center nucleus, where the potential is known to exhibit large and rapidly changing values.
\begin{figure}[htbp!]
    \centering
    \includegraphics[scale=0.5]{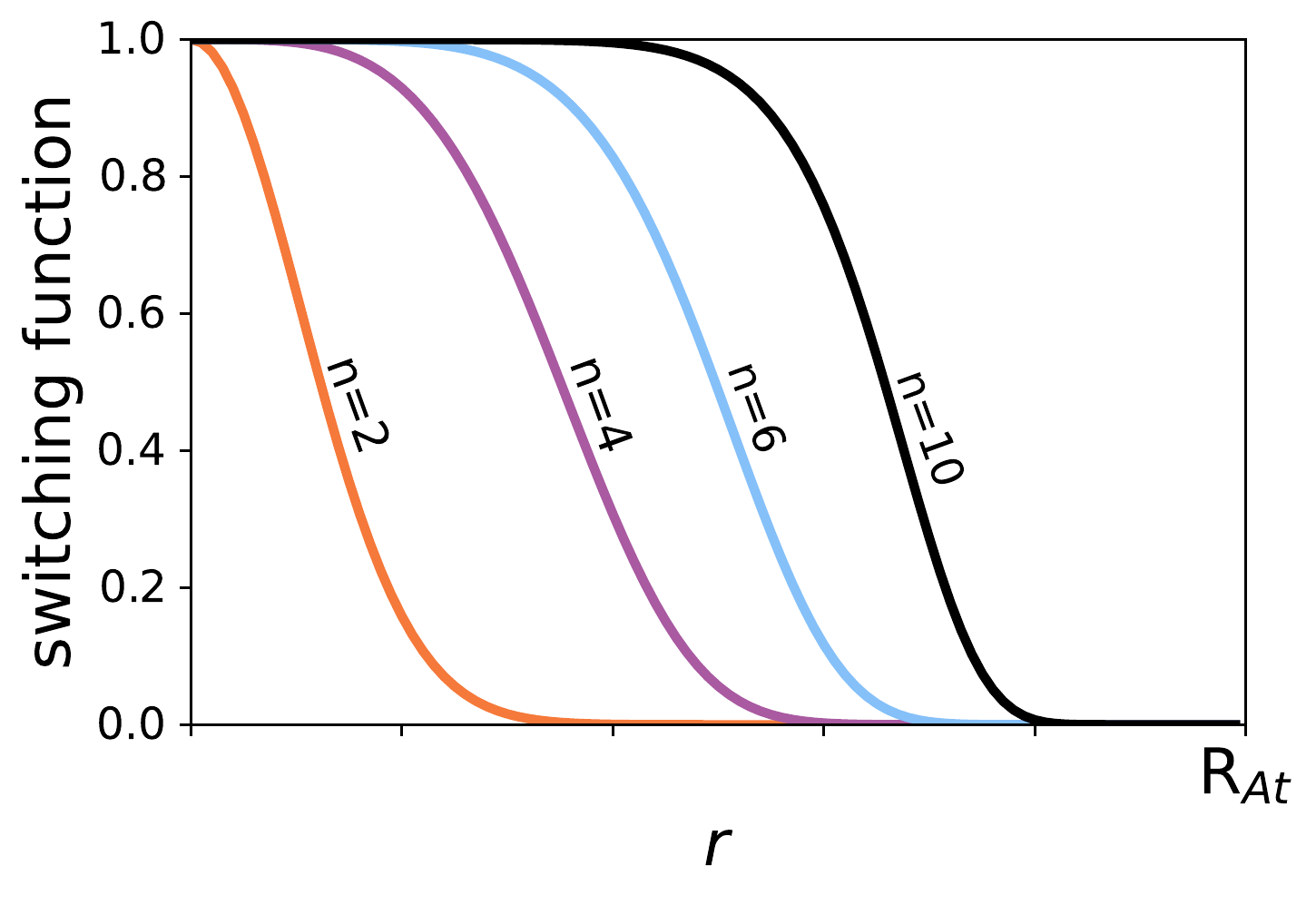}
    \caption{Switching cutoff function, Eq.\eqref{eq:fAT2},  shown for a range of powers $n$ where $R_{At}$ is the limit of the atomic grid. }
    \label{fig:powers_fitting}
\end{figure}
This problem is solved using Becke scheme, since both the source term of the Poisson equation and the integrand of the electrostatic potential can be partitioned in atomic components,
\begin{equation}
\begin{split}
\nabla^2\Phi_{1,a}&= -4\pi\,w^B_{a}\rho_1,\\
[\rho_1|\rho_2] &= \sum_{a}\int \mathrm{d}^3 r\,\,  \Phi_{1,a}\,w^B_a\,\rho_2+\\
&+\sum_{a,b\neq a}\int \mathrm{d}^3 r\,\,  \Phi_{1,a}\,w^B_b\,\rho_2.
\end{split}
\end{equation}
Here, $w^B_{a}$ are the Becke weights defined for atom $a$.
The charge distribution $w^B_{a}\,\rho_1$ and the corresponding potential $\Phi_{1,a}$ vary rapidly only near $\vec{R}_{a}$. The solution of the Poisson equation, therefore, can be accurately sought after in a reference frame centered on $\vec{R}_a$, e.g., by expanding it in a basis of auxiliary spherical functions. The second integration concerns the energy of a charge distribution close to a different center than the one from which the electrostatic potential originates. The potential is smooth at all points where the charge density is finite. This last component, therefore, is integrated to high precision using the atomic grid centered on $b$.

Since the weight functions $w^B_a(\vec{r})$ do not normally vanish asymptotically, the sphere within which the Poisson equation must be solved, or the integral evaluated, is as large as that of a single-center expansion. Once again, while this approach does eliminate the largest source of inaccuracy of the single-center expansion, it also multiplies the number of integration grids and with it the number of grid points needed to reach convergence. More importantly, this partitioning scheme fails to take advantage of the multipolar character of the interaction between well separated charges, requiring the solution of the Poisson equation over the full configuration space for each atom in the molecule. These two aspects are severe limitations in scattering calculations, where the wavefunction must be evaluated at radial distances significantly larger than the molecule itself. As a consequence, this approach can be predicted to scale poorly with the number of atoms in the molecule, as well as with the diameter of the molecular region, which can easily reach large dimensions even for few atoms, in the case of molecular dissociation~\cite{Laqua2018} or loosely bound van-der-Waals complexes. 

To treat integrals between basis functions that do not decay asymptotically, such as those involving continuum functions, it is essential for the atomic weights to decay rapidly, and be complemented by a single master weight and grid at large distances. The focus of the present study is to modify the Becke scheme to address the issues that arise when continuum functions are present, namely: i) its slow convergence, due to the mismatch between grids' and integration domain; ii) its redundancy, due to most of the points in the atomic grids that encompass the whole integration domain having negligible weight; iii) its inability to give rise to a natural partitioning of two-body interactions in short-range terms and long-range multipolar terms. This is accomplished by switching off the atomic weights $w_a(\vec{r})$ with an attenuation function that decreases with the distance from the molecular region. In addition, we introduce a single-center master grid, which accounts for those parts of configuration space outside the atomic centers and whose spherical boundary is ideally suited to enforce the asymptotics of scattering functions. 

We begin by modifying the Becke weight function, $w^B_a$ as,
\begin{equation}
\label{eq:modified-weights}
w^{At}_a(\vec{r}) = w^B_a(\vec{r})\, f^{At}(|\vec{r}-\vec{R}_{a}|), 
\end{equation}
where $f^{At}$ is a positive bell-shaped, monotonically decreasing radial function that starts from $1$ at the origin, with several zero derivatives, and becomes negligible beyond a given radius $R_{At}$:
\begin{equation}
\begin{split}
&f^{At}(x)\geq 0, \quad f(0)=1,\\
&(f^{At})^{(n)}(0)=0,\quad n<n_{max},\\
&f^{At}(x)\simeq 0\quad\forall\,\, x > R_{At}.
\label{eq:fAT}
\end{split}
\end{equation} 
One such possible function is
\begin{equation}
\begin{split}
&f^{At}(x;\alpha,n)=1-(1-e^{-\alpha x^2})^{n},\\
&\mathrm{for}\,\,  x\ll \alpha^{-1/2},\,\,f^{At}(x;\alpha,n)\sim 1 -\alpha^n x^{2n} + o(x^{4n}),\\
&\mathrm{for}\,\,  x\gg \alpha^{-1/2},\,\,f^{At}(x;\alpha,n)\sim n\,e^{-\alpha x^2}.
\label{eq:fAT2}
\end{split}
\end{equation} 
Here \(x=|r-R_a|/R_{At}\), where \(R_{At}\) is the radius of the atomic grid associated with atomic nucleus at \(R_a\). The value of $\alpha$ such that $w^{At}_a(\vec{r})$ is smaller than a certain threshold $\delta$ beyond a given radius $R_{At}$ may be computed using,
\begin{equation}
\label{eq:scheme1Parameters}
f^{At}(R_{At}) = \delta \quad \implies\quad \alpha = R_{At}^{-2}\ln(n/\delta).
\end{equation}
Examples of such functions for different order $n$ are shown in Fig. \ref{fig:powers_fitting}.
The introduction of the complementary weight function, 
\begin{equation}
\label{eq:central-grid-weights}
 w^{At}_0(\vec{r})=1-\sum_a w^{At}_a(\vec{r}),
 \end{equation}
restores the partition of unity with the desired properties of the grid at long range, 
\begin{equation}\label{eq:ModifiedBeckesWeights}
\sum_{a=0,N} w^{At}_a(\vec{r})=1,\quad \lim_{r\to\infty}w^{At}_{a>0}(\vec{r})=0,\quad \lim_{r\to\infty}w^{At}_{0}(\vec{r})=1.
\end{equation}
The partitioning scheme that uses this partition of unity, $\{w^{At}_a\}_{a\in\{0,1,2,\ldots,N\}}$, will be denoted scheme 1.
Alternatively (scheme 2), one can use a single switch-off function $f^{Mol}(\vec{r})$ for the whole molecule,
\begin{equation}\label{eq:molwt}
w^{Mol}_a(\vec{r}) = w^B_a(\vec{r}) f^{Mol}(\vec{r}).
\end{equation}
The switch-off function $f^{Mol}(\vec{r})$ can be easily defined from any scalar field $V(\vec{r})$ that diverges at all nuclei and vanishes at large distances, such as
\begin{equation}
f^{Mol}(\vec{r})=f^{Mol}(\vec{r};\alpha,n) = f^{At}\left(N/V(\vec{r});\alpha,n\right),
\end{equation}
with the electrostatic potential due to singly-charged nuclei being a possible choice,
\begin{equation}
\label{eq:weight_pot3}
V(\vec{r}) = \sum_{a=1}^N |\vec{r}-\vec{R}_a|^{-1}.
\end{equation}
In this case, $w_0$ coincides with $1-f^{Mol}(\vec{r};\alpha,n)$.
In Sec.~\ref{sec:3}, we examine the performance of the partition of unity based on both the atomic (1) and the molecular (2) schemes.

Both approaches have manifest advantages over the original Becke scheme. First, all integrals over a spherical volume are much more accurate, since the quadrature grid matches the integration domain, whereas in the original Becke scheme the polycentric atomic grids contain but do not coincide with the integration domain, resulting in a first-order integration scheme, where the error is inversely proportional to the number of integration points, for any integrand that does not vanish smoothly before reaching the boundary of the integration volume. Second, in the Poisson equation for any $a>0$,
\begin{equation}
\nabla^2\Phi_{1,a}(\vec{r})  = -4\pi\rho_1(\vec{r})w^{At}_{a}(\vec{r}),
\end{equation}
the term on the RHS is essentially zero beyond the finite radius $R_{At}$ of any atom. As a consequence, the solution for $r>R_{At}$ is conveniently expressed in terms of a multipolar expansion, which is analytic and easy to evaluate,
\begin{subequations}
\begin{equation}
\Phi_{1,a}(\vec{r}) = \sum_{\ell m} \frac{M_{\ell m;1,a}}{r^{\ell+1}} Y_{\ell m}(\hat{r})
\end{equation}
\begin{equation}
M_{\ell m;1,a} =\frac{4\pi}{2\ell+1}\int  r^{\ell}Y_{\ell m}^*(\hat{r})w_a(\vec{r}) \rho_1(\vec{r})\,\mathrm{d}^3r.
\end{equation}
\end{subequations}
This approach significantly reduces the size of the sphere around each atom, and with it the region of space where the Poisson equation has to be solved numerically and is sensitive to the distribution of the grid points. 


 \section{Numerical results}
\label{sec:3}
This section is devoted to compare the performance of the modified Becke schemes 1 and 2 with the original Becke approach and with a single center spherical grid for a set of integrals representative of those encountered in a scattering or photoionization calculation. We consider integrands that exponentially decrease at large distances, such as nuclear centered Gaussians, that oscillate to large radial distances, such as spherical Bessel functions of varying energy located on the central grid, as well as functions exhibiting singularities at the atomic nuclei.  In the latter case, we consider integrals involving multicenter Yukawa potentials, where the result is known analytically for very large integration volumes, and which are analogous to nuclear-attraction integrals.
The application of the modified Becke methods to bielectronic integrals and to the calculation of physical observables such as molecular photoionization cross sections is beyond the scope of the present work and will be examined elsewhere. 


We carry out the numerical tests using as a reference the nuclei of the NO$_2$ molecule in its equilibrium configuration in the $yz$ plane, as specified in 
Table~\ref{tab:coords}. 

\begin{table}[hbtp!]
\caption{\label{tab:coords} Atom Coordinates (a.u.)}
\begin{tabular}{ccc}
\hline \hline
Atom & $y$ & $z$   \\
 \hline
N  & \phantom{-}0 & \phantom{-}0.6909 \\
O1  & \phantom{-}2.1403  & -0.3022  \\
O2  & -2.1403  & -0.3022  \\ 
 \hline\hline
\end{tabular}
\end{table} 

Every atomic integration region is subdivided into a set of spherical shells. The 3D grid points within a given shell are a product of a Gauss-Legendre radial and Lebedev angular~\cite{Lebedev1975, Lebedev1976, Lebedev1977, Lebedev1992, Lebedev1995, Lebedev1999} set of points. Different shells have have the same number of angular points, but can differ in the number of radial points. 
\begin{figure*}[hbtp!]
\centering
\includegraphics[width=0.85\textwidth]{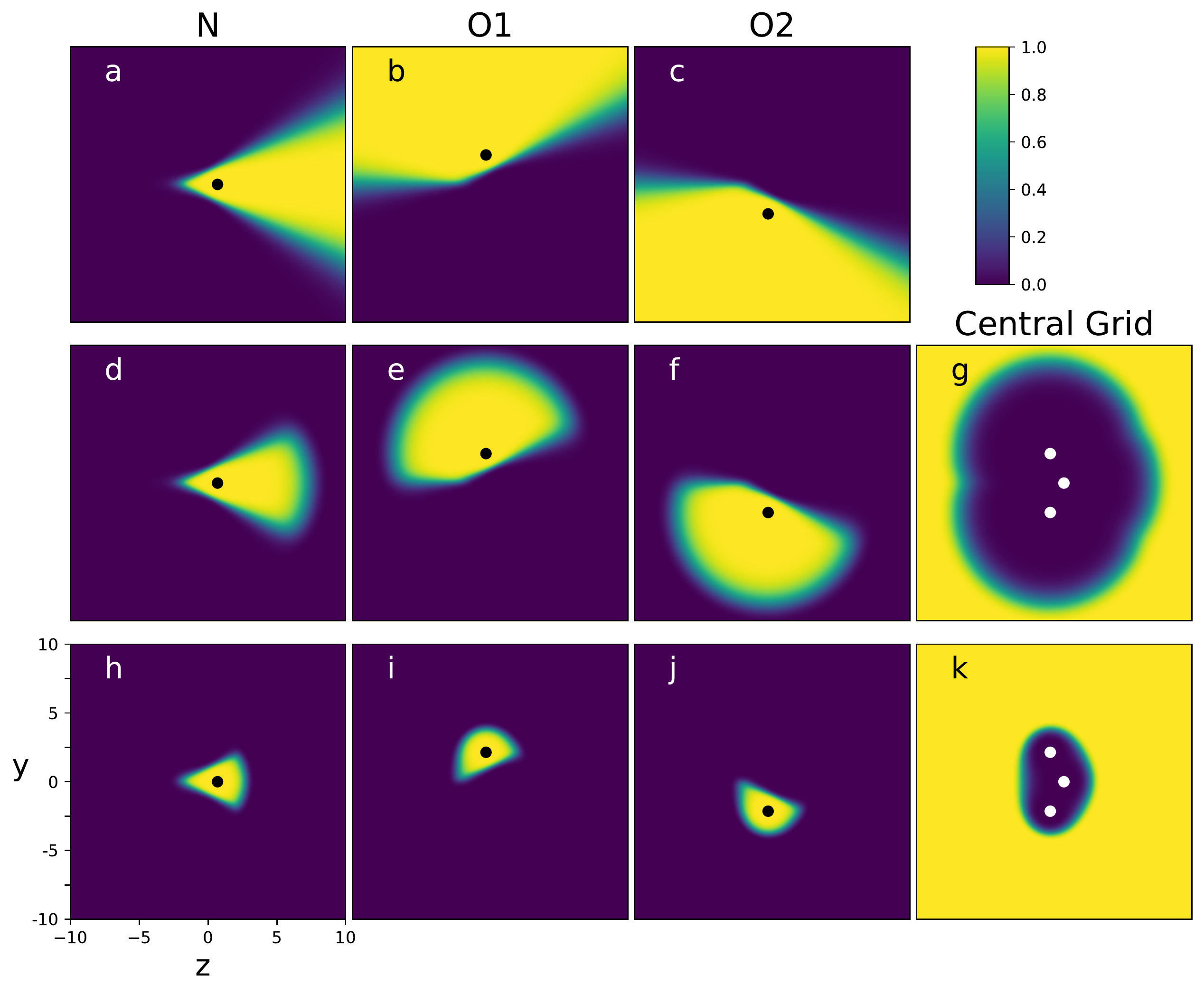}
\caption{Original and modified Becke’s weight distributions for NO$_2$ molecule placed at coordinate points of Table \ref{tab:coords}. Atomic grid sizes are $R=10$~a.u. The original Becke scheme defines atomic weights that in general do not vanish at infinity (a-c). The atomic based modified weights (scheme 1) and the 
molecular based weights (scheme 2) are given in (d-g) and (h-k), respectively for the same atomic parameters. Panes (g) and (k) give the complementary weights for the central grid in the two modified schemes. The solid circles represent the location of atomic nuclei.}
    \label{fig:triatomic_becke}
\end{figure*}

Fig.~\ref{fig:triatomic_becke} shows the range of the original and modified weight functions in NO$_2$. Whereas the weights in the modified schemes 1 and 2 (middle and lower panels, respectively) rapidly drop to zero as the distance from the their atomic centers of reference increases, the original Becke weights $w_a^B(\vec r)$ (upper panels) do not vanish asymptotically at all.   
For integrands that do not decrease rapidly outside the molecular region, such as those involving Rydberg and scattering orbitals, therefore, the original Becke scheme requires disproportionately large integration grids for each and every atom. In the modified Becke schemes, the atomic weights smoothly approach zero at the atomic shell boundaries, set at a fixed distance $R$ from each nucleus. 
As a consequence, the new atomic weights maintain their ability to describe compact functions near the nuclei, while the central grid captures integrands of long-range components. 
In scheme 1, the complementary weight $w_0^{At}$ (panel g, Fig.~\ref{fig:triatomic_becke}) exhibits comparatively sharp features at the intersection between the boundary support of atomic weights, thus affecting the convergence of the integral as a function of the angular points in the central grid. This problem is less prominent for larger atomic spheres but it is exacerbated if the atomic weights are strongly confined.
The bottom panels in Fig.~\ref{fig:triatomic_becke} show the three atomic (h-j) and the complementary weight (k) for scheme 2. Even if the parameters are chosen to yield more localized atomic weights than in scheme 1, the hole profile in the complementary weight is considerably smoother, which suggests a smaller number of angular points is needed to reach convergence, if strongly localized atomic weights are required.

Let us begin by considering  $s$-type GTOs using the same grid on each atom, defined with $N_r$ number of radial and $N_{\Omega}$ number of angular grid points, for a total number of points $N_G=N_R \cdot N_\Omega$.
The number of angular points $N_{\Omega}$ is dictated by the maximum angular momentum $\ell_{\rm max}$ for which the Lebedev quadrature is exact. 
The density of radial points of the atomic shell is larger near the nucleus, to achieve higher accuracy for the compact functions. The GTOs are placed at each atomic center with different exponential factors $\alpha$ to cover the cases from compact to diffuse orbitals. For each scheme, we set a convergence tolerance for all GTOs of at least 10$^{-10}\%$. 

In the single-center expansion, the radial grids extends to $R_c=80$ a.u. to cover the most diffuse GTOs, and it has a smaller radial spacing in the molecular region. A Lebedev quadrature of $\ell_{\rm max}=131$ is necessary to maximize the integration accuracy for the compact GTOs. The results, in Table \ref{table:GTO-1}, show that while the single-center expansion can accurately describe the GTOs on the nitrogen atom, it fails to represent GTOs with exponent $\alpha\ge200$ on the oxigen atoms, which are furher away from the expansion center. Increasing the number of radial points does not improve the accuracy of the most compact GTOs. 
Additional tests using products of $\theta$ and $\varphi$ grids, with up to ten times more points than the Lebedev grids used in Table \ref{table:GTO-1}, still have errors larger than 10$^{-5}$ for the compact Gaussians. 

In the original Becke scheme, with atomic shells extending to $R=80$ a.u., Table \ref{table:GTO-1} shows that quite accurate results can be achieved with an $\ell_{\rm max}=83$. The contrast with the single-center expansion is particularly dramatic for the GTOs centered on the oxygen atoms, since the integration accuracy improves by eight orders of magnitude.  
In general, all of the Becke schemes perform significantly better than the single-center expansion, and the modified Becke scheme reach high accuracy with fewer points than the original Becke scheme. In addition, the modified Becke scheme does not require a high density of radial points near the origin of the central grid.
\begin{table}[b]
    \centering
    \caption{\label{table:GTO-1} Relative error between numerical and analytical results for integration of Gaussian functions over a range of GTO exponents, $\alpha$, using a single-center grid (SC), the original Becke scheme and the modified Beckes's scheme with the associated total number of grid points, N$_{\rm G}$, (see text for details). The notation $[n]$ stands for $\times 10^{n}$.}
\begin{tabular}{c|r|l|l|l}
\hline \hline
 Atom & $\alpha$ & SC& Becke& Mod. Becke\\
\hline
\multirow{8}{*}{N} 
 &\,\,1000  &\,\,   1.76 [-13]  &\,\, 4.97 [-11] &\,\, 5.38 [-11] \\
 &\,\, 500  &\,\,   2.52 [-14]  &\,\, 2.55 [-11] &\,\, 9.82 [-11] \\
 &\,\, 200  &\,\,   4.61 [-14]  &\,\, 4.71 [-11] &\,\, 7.92 [-11] \\
 &\,\, 100  &\,\,   5.83 [-14]  &\,\, 3.03 [-13] &\,\, 4.09 [-11] \\
 &\,\,  10  &\,\,   1.72 [-13]  &\,\, 1.96 [-13] &\,\, 1.33 [-13] \\
 &\,\,   1  &\,\,   1.41 [-13]  &\,\, 1.72 [-13] &\,\, 7.29 [-14] \\
 &\,\, 0.1  &\,\,   2.08 [-14]  &\,\, 1.79 [-13] &\,\, 4.43 [-14] \\
 &\,\,0.01  &\,\,   4.97 [-13]  &\,\, 2.76 [-13] &\,\, 2.59 [-13] \\
 \hline
\multirow{8}{*}{O} 
 &\,\, 1000  &\,\,   3.72  [-2] &\,\, 4.91 [-11] &\,\, 4.22 [-11]\\
 &\,\,   500 &\,\,   8.05  [-4] &\,\, 1.80 [-11] &\,\, 4.52 [-11]\\
 &\,\,   200 &\,\,   1.26  [-6] &\,\, 5.39 [-11] &\,\, 2.78 [-11]\\
 &\,\,   100 &\,\,   2.80 [-12] &\,\, 3.32 [-11] &\,\, 4.14 [-11]\\
 &\,\,    10 &\,\,   7.88 [-14] &\,\, 2.27 [-13] &\,\, 7.28 [-13]\\
 &\,\,     1 &\,\,   1.23 [-13] &\,\, 7.97 [-14] &\,\, 4.07 [-14]\\
 &\,\,   0.1 &\,\,   7.38 [-14] &\,\, 1.90 [-13] &\,\, 6.49 [-14]\\
 &\,\,  0.01 &\,\,   4.82 [-13] &\,\, 2.55 [-13] &\,\, 2.67 [-13]\\
\hline  
$N_G$ &\,\, &\,\, 1,801,100 &\,\, 2,602,500&\,\,1,790,440\\
$\ell_{\rm max}$ &\,\, &\,\, 131 &\,\, 83&\,\,83\\
\hline \hline
\end{tabular}
\end{table}

\begin{figure*}
    \centering
    \includegraphics[width=0.85\textwidth]{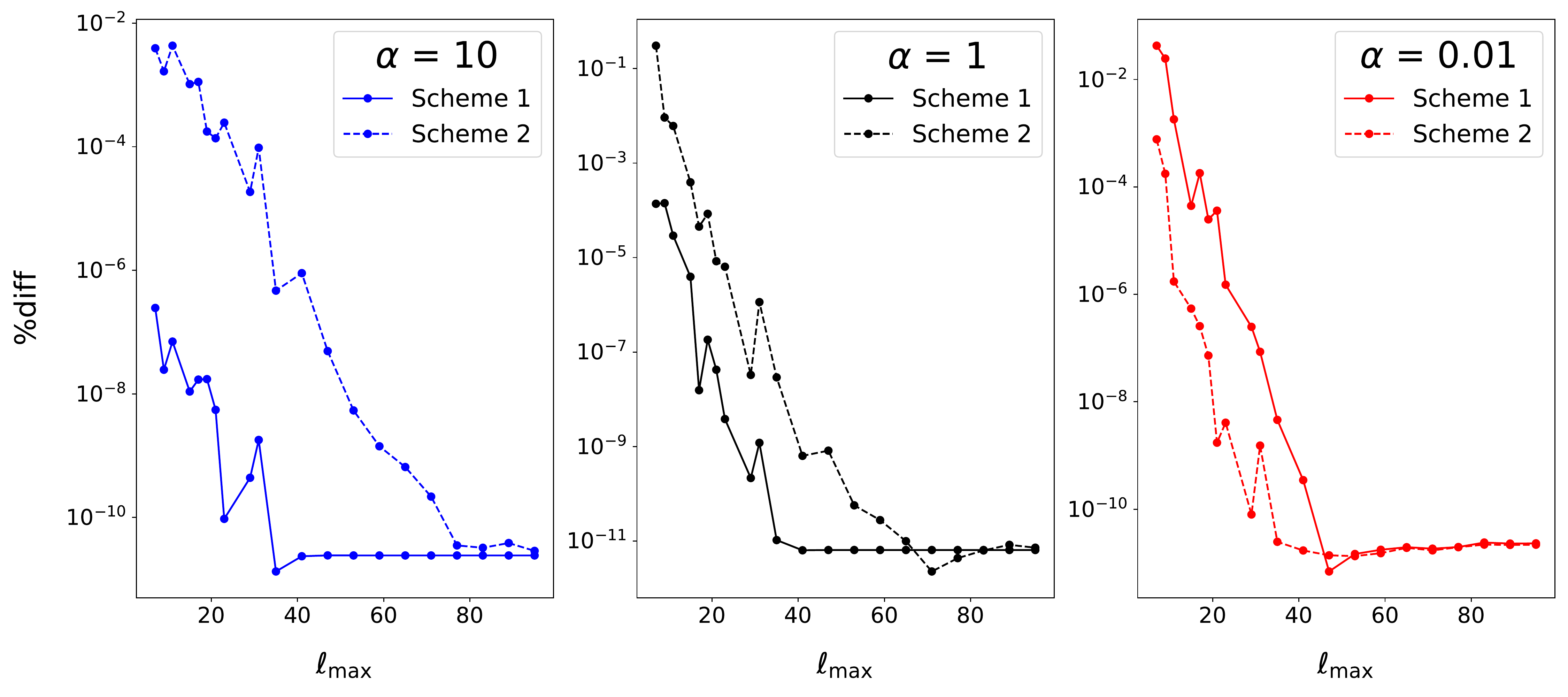}
    \caption{\label{fig:convergence_modbecke} Modified-Becke's accuracy convergence for scheme 1 vs. scheme 2 as a function of angular momentum number $\ell_{\rm max}$ of the central grid. The test is for an O atom in the NO$_2$ molecule placed at coordinate points of Table \ref{tab:coords}. Atomic grid sizes are $R=10~$a.u. set in a central grid of of size $R=80~$a.u.}
\end{figure*}

Fig.~\ref{fig:convergence_modbecke} compares the two modified Becke schemes for three different Gaussians centered on the O atom.  Since the modified Becke schemes contain a central grid, one can considerably shrink the sizes of the atomic radii, here chosen as $R=10$a.u.  For the more compact Gaussians, the modified Becke scheme I does better than scheme II, whereas the opposite is true for the diffuse Gaussians. 
Overall, however, scheme I shows better control of the error over a broader range of exponential parameters. 
In the rest of the article, therefore, we will focus on Scheme I.

It is instructive to analyze the dependence of the modified Becke scheme on the size of the atomic shells,  where the transition from the near-nuclei Becke scheme to a single-center grid method occurs. Using different shells with radii $R=2.5$, $5.0$, $10.0$ and $20.0$ a.u., the requested accuracy for all GTOs was always reached with the modified scheme with less than 2 million points. 
As the radius $R$ decreases, the density of grid points at small distances must be increased for both the atomic and the complementary grid, to efficiently (see panel (d) of Fig. \ref{fig:triatomic_becke}) represent the variation of the switch-off function $f^{At}$. 
In fact, using a smaller radius $R$ allowed us to reach higher accuracy ($\sim10^{-11}$ instead of $\sim10^{-9}$) for the most compact GTOs when compared with the case $R=10$~a.u. shown in Table \ref{table:GTO-1}.
In general, however, the accuracy is excellent for all shell radii.
   
Let us now consider the accuracy of the different weighting schemes for evaluating integrals of the Yukawa potential,
\begin{equation}
V_b(\vec r)= \frac{e^{-|\vec r-\vec R_b|}}{|\vec r-\vec R_b|}.
\end{equation}
This test is more demanding than Gaussian integrals due to the Coulomb singularity at the atomic centers.  The results are shown in Table~\ref{tab:yukawa}, where we used the same parameters as previously for the Gaussian tests. Both the original and the modified Becke schemes are superior in accuracy by many orders of magnitude than the single-center grid method, and, as expected for a fast-decreasing integrands such as the Yukawa potential, the modified Becke scheme is comparable in accuracy to the original. 

\begin{table}[b]
\caption{\label{tab:yukawa} Relative error between numerical and analytic results for the integration of Yukawa integrals on N and O atoms for the three different schemes.}
\begin{tabular}{cccc}
\hline \hline
 Chem. Symb. & Single-Center &  Becke & Mod. Becke \\
\hline
\multirow{1}{*}{N}  & 1.65 [-5] &   2.90 [-13] & 1.99 [-14]\\
 \hline
\multirow{1}{*}{O} & 2.01 [-4] &   3.25 [-13] & 1.97 [-14]\\
\hline \hline
\end{tabular}
\end{table}

So far, we have shown the equivalence of the original and the modified Becke scheme using functions that decay at large radial distances. The goal, however, is to develop a weighting scheme that is effective not only for the short-range functions examined in previous sections, but also for the continuum functions found in scattering and photoionization calculations, such as spherical Bessel functions, B-splines or finite element DVR functions. Photoelectrons can easily have de Broglie wavelengths of the order of one atomic unit, i.e., comparable to the width of the switch-off function used in the modified Becke schemes, and they rapidly oscillate over large (indeed, infinite) radial distances. 
A few examples illustrate the superior performance of the modified Becke scheme when continuum functions are involved. In the first test, we compute the overlap of the spherical Bessel function $j_0(kr)=\sin(kr)/kr$ with various Gaussian functions, for Gaussian exponents between $.002$ and $.005$ and radial photoelectron momentum $k$ varying from $1$eV to $400$eV (1a.u.$\simeq$27.21 eV).  The atomic spheres were chosen to have a radius of $R=5$ a.u..  While the original Becke scheme can compute these integrals, the modified scheme could do so with a reduction of $60$ to $90$\% in the number of points for all energies. These results illustrate the efficiency of the master grid to accurately capture the behavior of these integrands at large radial distances.

For the second test, shown in Table~\ref{table:sphericalBessels}, we compute the integral over a spherical region of the two functions $j_0(r)$, and $r j_1(r)$ where $j_\ell(r)$ are spherical Bessel functions.  Our reason for testing the integral of $rj_1(r)$ rather than $j_1(r)$ itself, is  that $j_1(r)$ has a discontinuous radial derivative at the origin, something that would never occur in practice using a set of full 3D basis functions.
The dramatic improvement in accuracy of the modified vs the original Becke method, is clear.  The poor performance of the original Becke scheme is due to the mismatch between the boundary of the integration region, which is dictated by the physics of the problem, and the boundary of off-center spherical quadrature grids. Apart from the larger number of points required by the Becke scheme, the method simply will not be usable for the computation of the integrals needed in scattering and photoionization processes with general numerical bases.
\begin{table}[]
\caption{\label{table:sphericalBessels} Relative error between numerical and analytic results for the integration of spherical Bessel integrals for the Becke and modified Becke schemes.}
\begin{tabular}{ccccccc}
\hline \hline
function                                    & \multicolumn{1}{|c|}{Method} & \multicolumn{1}{c|}{\begin{tabular}[c]{@{}c@{}}Total \\ radial points\end{tabular}} & \multicolumn{1}{c|}{\begin{tabular}[c]{@{}c@{}}angular points\\ per radial shell\end{tabular}} & \multicolumn{1}{c|}{\begin{tabular}[c]{@{}c@{}}Total num. \\ grid points\end{tabular}} & \multicolumn{1}{c|}{\begin{tabular}[c]{@{}c@{}}Total points inside\\  integration region\end{tabular}} & \begin{tabular}[c]{@{}c@{}}\%diff with\\ analytical\end{tabular} \\ \hline 
\multicolumn{1}{c|}{\multirow{2}{*}{$j_0$}} & Becke                       & 1000                                                                                & 974                                                                                            & 974000                                                                                 & 769744                                                                                                 & 4.11{[}-2{]}                                                     \\ \cline{2-7} 
\multicolumn{1}{l|}{}                       & Mod. Becke                  & 900                                                                                 & 974                                                                                            & 876600                                                                                 & 876600                                                                                                 & 3.52{[}-12{]}                                                    \\ \hline
\multicolumn{1}{c|}{\multirow{2}{*}{$r \cdot j_1$}} & Becke                       & 2000                                                                                & 974                                                                                            & 1948000                                                                                & 1370186                                                                                                & 1.16{[}-2{]}                                                     \\ \cline{2-7} 
\multicolumn{1}{l|}{}                       & Mod. Becke                  & 1800                                                                                & 974                                                                                            & 1753200                                                                                & 1753200                                                                                                & 1.92{[}-11{]}                                                     \\ \hline \hline
\end{tabular}
\end{table}

\section{Conclusions and Outlook} \label{sec:4}
In this paper, we have developed a new variant of the popular Becke molecular integration method which is able to efficiently integrate functions containing integrands that do not vanish asymptotically, such as those appearing in typical scattering and photoionization calculations. This required a modification of the Becke weighting to confine the grid points on the atomic sites to molecular dimensions and to augment the atomic grid with a single, master grid to capture the interstitial and long range behavior of the integrand. The new approach has a number of appealing properties that should be stressed. In addition to its high accuracy for integrands which do not decay rapidly away from the atomic centers, it provides a very efficient approach to solving the Poisson equation and ultimately to the calculation of the hybrid one-and-two electron integrals encountered in scattering calculations.

The numerical study confirmed that single-center grids are incapable of providing very accurate  integrations with integrands having strongly localized integrands close to the atomic centers. Integration approaches which partition space into atomic sub-regions are far superior, as already shown by Becke.  By weighting each atomic region, the molecular integral may be decomposed into a sum of atomic-like contributions, which can be easily discretized to produce an accurate answer.  For integrands which vanish sufficiently rapidly at large distances, the original weighting method of Becke which partitions the atomic regions into ``fuzzy", overlapping regions, works remarkably well. For integrals that extend over large regions of space, such as those that appear in scattering calculations, however, the  Becke scheme becomes unnecessarily expensive, as a consequence of the atomic Becke weights having a non-vanishing amplitude at large distances, and inaccurate, since the boundary of the integration grids do not match the boundary of a single spherical integration domain.  We have presented two extensions of the Becke method that confine the atomic weights to the molecular region, completes them with a complementary central grid and weight and accurately and economically captures the large-radius region, where the charge densities are regular. The new weighting schemes produce quite accurate results for test functions such as compact Gaussian functions, Yukawa potentials, and spherical Bessel functions, which are representative of the functions encountered in realistic molecular scattering problems.  The method scales favorably with the number of atoms in the molecule and with the size of the integration volume, when compared with the original Becke scheme, as shown in the Appendix. Furthermore, it is expected to yield uniformly accurate results even for dissociating molecules.

The efficiency of the modified schemes compared to the original Becke scheme increases rapidly with the size of the molecule as a consequence of more localized nature of the weighted charge distributions, allowing simple multipole methods to be used away from the atomic sites. For one-body integrals, the efficiency increases linearly with the number of atoms.  In Appendix~\ref{sec:compest}, we provide informed estimates of the computational cost for both the Becke and Modified Becke methods in calculating the two-electron integrals.  The estimates show that the modified Becke method can lead to a several order of magnitude speedup over the Becke method when applied to these integrals.  

\section{Acknowledgments}
The work of N.D. and L.A.~was supported by the United States National Science Foundation under NSF grant No.~PHY-1607588, by the DOE CAREER grant No.~DE-SC0020311 and the work of B.I.S and H.G. was supported by the Department of Commerce, National Institute of Standards and Technology.

\appendix
\section{Computational Cost}
\label{sec:compest}
This appendix estimates and compares the computational costs of evaluating one-electron and two-electron integrals in the Becke and Modified Becke schemes. Table~\ref{table:defs} defines some parameters employed in this analysis. 
\begin{table}[h]
\caption{\label{table:defs} Parameters relevant to the cost of the Becke ($b$) and Modified Becke ($mb$) integrations.}
\begin{tabular}{c|l|l}
\hline \hline
 Variable &  Definition  & Value\\
\hline \hline
$N_{at}$    &    Number of atoms & \\
\hline
$R_{cg}$    & Radius of central grid            & $25-50$~a.u.  \\
\hline
$R_{at}$    & Radius of atom                    & $5-10$~a.u.  \\
\hline
$r_g$       & Ratio of radii                    & $R_{at}/R_{cg}\lesssim 0.2$  \\
\hline
$N_{cg}$    & Number of points in central grid  &                             \\
\hline
$N_{ag}^{b}$& Number of atomic points in        &                             \\
                                  & Becke scheme& $\approx N_{cg}$ \\
\hline
$N_{ag}^{mb}$& Number of atomic points in       &                              \\
            & modified Becke scheme             & $ N_{cg} r_g^3 \approx N_{cg}/125$  \\
\hline

$N^b$       & Total number of Becke points      & $\approx N_{cg}  N_{at}$  \\
\hline
$N^{mb}$    & Total number of modified          &  $N_{cg}+ N^{mb}_{ag} N_{at}$ \\
            & Becke points                      &  =  $ N_{cg}[ 1 +  N_{at}\,r_g{^3}] $ \\
\hline
$N_{op}$    & flops to evaluate auxiliary functions& \\
            & in the Poisson Equation solution \\
\hline
$N_{nn}$    & Number of nearest neighbor atoms  &  \\
\hline \hline
\end{tabular}
\end{table}
In the definition of $N^b$, we have made the reasonable assumption that, in the Becke scheme, we will need the same number of points per atom as is needed for the central grid in the modified Becke scheme. Indeed, each atomic grid in the Becke scheme needs to cover the entire computational volume. In the modified Becke scheme, on the other hand, only the central grid covers the entire computational volume, while the individual atomic grids are restricted to an atomic radius. We also assume that the number of grid points scales proportionally to the volume of the enclosed computational region.

The size of the central grid is dictated by the asymptotics of a scattering problem and needs to be sufficiently large that one may neglect interchannel coupling. Here, we assume $r_g=0.2$ as a reasonable upper bound to the ratio of the atomic and central grid radii. As a consequence, in the evaluation of single-electron integrals, the computational cost of the Modified Becke approach is dominated by the sum over the points of the master grid, which is equivalent to the sum over a single atomic grid in the original Becke approach. For one-electron integrals, therefore, the Modified Becke scheme is more efficient than the original Becke scheme by a factor comparable to the number of atoms in the system.

If the reduction of the computational cost of one-electron integrals is already appreciable, the gain for two-electron integrals is dramatic. In the Modified Becke scheme, a two-electron integral $[\rho|\rho']$ is evaluated by partitioning the densities as follows
\begin{equation}
[\rho|\rho']=[\rho_0|\rho'_0] + \sum_{a}\left( [\rho_a|\rho'_0] + [\rho_a'|\rho_0] \right)
+\sum_{a,b\in I^{NN}_{a}} [\rho_a|\rho'_b] +\sum_{a,b\notin I^{NN}_a} [\rho_a|\rho'_b],
\end{equation}
where $\rho_0 = w_0\rho$, $\rho_a = w_a\rho$, etc. [compare with \eqref{eq:ModifiedBeckesWeights}] and $I^{NN}_a$ is the set of ``nearest-neighbor" atomic indices whose integration spheres intersect that of atom $a$, including $a$ itself. 
The evaluation of each term $[\rho_a|\rho'_b]$ requires two steps: i) the determination of the potential $V_a$, originating from $\rho_a$, and ii) the integration of $V_a\rho_b$.  Evaluating the potential $V_a(\vec{r})$ at points outside the support of $\rho_a$ is inexpensive, since the potential can be expanded in multipoles. To evaluate the potential within the support of $\rho_a$, on the other hand, it is necessary to solve the Poisson equation (PE) by the inverting a discretized radial Laplace operator and applying it to every spherical density component, $\rho_{a,\ell m}(r)$, where $\rho_a(\vec{r}) = \sum_{\ell m}\rho_{a,\ell m}(r) Y_{\ell m}$.  Finally, step (2) requires evaluating $V_a$ at the points in sphere $b$ and then carrying out the final integration by summing over all points in sphere $b$. 
In the original Becke scheme, all atomic grids overlap and all have the same large size. In modified Becke, only one grid is large, and all the other interactions are either between few overlapping spheres, one of which is small, or between many disjoint small spheres. Most of the interaction, therefore, is multipolar. 
The computation cost for each step for both schemes is listed in Table~\ref{table:2el}, where $N_{nn} = N_{at}^{-1}\sum_a o(I^{NN}_a)$ is the average number of nearest neighbors. Here, $o(I)$ stands for the number of elements of set of indexes $I$.
Using the value of $N^{mb}_{ag}=N_{cg}\,r_g^3$ given in Table~\ref{table:defs}, the number of operations in the modified Becke method, $N_{\mathrm{flop}}^{mb}$, compared to Becke, is
\begin{equation}
\label{eq:ops}
\frac{N_{\mathrm{flop}}^{mb}}{N_{\mathrm{flop}}^{b}}\approx \frac{N_{nn}}{N_{at}}\Big[ r_{g}{^{6}} + \frac{r_{g}{^{3}}}{N_{nn}(N^{mb}_{ag})^{1/3}}.
\Big] 
\end{equation}
Since $r_g\lesssim 0.2$, the Modified Becke scheme requires three to five orders of magnitude less operations than the original Becke scheme. 
 
\begin{table}[h]
\caption{\label{table:2el} Comparison of Computational Cost of Becke and Modified Becke Method.
The third row shows the individual contributions of the overlapping, non-overlapping and central grids. The number of radial and angular points in a spherical grid with $N$ points is estimated as approximately $N^{1/3}$ and $N^{2/3}$, respectively.}
\begin{tabular}{c|c|c}
\hline \hline
 & Becke &  Mod. Becke \\
\hline \hline
Tabulation of $\rho$ &      $N^b=N_{at}N_{cg}$ &  $N^{mb}=N_{cg}+N_{at}N^{mb}_{ag}$ \\
 \hline
Solution of $V_a$ &     $N_{at}[N^{b}_{ag}]{^{4/3}}\approx N_{at}[N_{cg}]{^{4/3}} $ & $N_{at} [N^{mb}_{ag}]^{4/3} + [N_{cg}]{^{4/3}}  \approx[N_{cg}]{^{4/3}} $ \\
 \hline
Evaluation of $V_a$ & $N_{at}[N_{at} - 1]/2  [N^{b}_{ag}]^2 N_{op}$ & $N_{nn}N_{at}[N^{mb}_{ag}]^{2} N_{op}$  \\
 & $\approx  N_{at}[N_{at} - 1]/2 [N_{cg}]^2N_{op}$ & + $N_{at}[N_{at}-N_{nn}-1][N^{mb}_{ag}]^{5/3}$     \\
 & & $ + N_{at}N_{cg}[N^{mb}_{ag}]^{2/3}$  \\
 \hline
Final Integration & $\sim N^b$ &  $\sim N^{mb}$ \\
\hline \hline
Total & $ \approx (N_{at}^2 N_{cg}^2 N_{op})$ & $\approx N_{nn}N_{at}[N^{mb}_{ag}]^{2} N_{op} + N_{at}N_{cg}[N^{mb}_{ag}]^{2/3}$ \\
\hline\hline
\end{tabular}
\end{table}

\bibliography{Manuscript.bib}

\end{document}